\documentclass[aps,preprint]{revtex4}
\usepackage[dvips]{graphicx}
\usepackage{psfig,epsfig}
\usepackage{amsmath}
\usepackage{amsfonts}
\usepackage{amssymb}
\usepackage{wasysym}
\usepackage{color}

\begin{document}
\def\be{\begin{equation}}
\def\ee{\end{equation}}
\def\bfi{\begin{figure}}
\def\efi{\end{figure}}
\def\bea{\begin{eqnarray}}
\def\eea{\end{eqnarray}}

\title{Test of Local Scale Invariance from the direct measurement of the response function in the 
Ising model quenched to and to below $T_C$}

\author{Eugenio Lippiello$^\ddag$, Federico Corberi$^\dag$
and Marco Zannetti$^\S$}
\affiliation {Istituto Nazionale di Fisica della Materia, Unit\`a
di Salerno \\ and Dipartimento di Fisica ``E.Caianiello'',
Universit\`a di Salerno,
84081 Baronissi (Salerno), Italy}

\ddag lippiello@sa.infn.it
\dag corberi@na.infn.it 
\S zannetti@na.infn.it

\begin{abstract}

In order to check on a recent suggestion that local scale invariance 
[M.Henkel et al. Phys.Rev.Lett. {\bf 87}, 265701 (2001)]
 might hold
when the dynamics is of Gaussian nature, we have carried out the
measurement of the response function in the kinetic Ising model with Glauber dynamics quenched to $T_C$ in $d=4$,
where Gaussian behavior is expected to apply, and in the two other cases of the $d=2$ model quenched to $T_C$ and
to below $T_C$, where instead deviations from Gaussian behavior are expected to appear.
We find that in the $d=4$ case there is an excellent agreement between the numerical data, the local
scale invariance prediction and the analytical Gaussian approximation. No logarithmic corrections are numerically 
detected. Conversely, in the $d=2$
cases, both in the quench to $T_C$ and to below $T_C$, sizable deviations of the local scale invariance
behavior from the numerical data are observed. These results do support the idea
that local scale invariance might miss to capture the non Gaussian features of the dynamics.
The considerable precision needed for the comparison has been achieved
through the use of a fast new algorithm for the measurement of the response function without applying the external field.
From these high quality data we obtain
$a=0.27 \pm 0.002$ for the scaling exponent of the response function in the $d=2$ Ising model quenched to below $T_C$, 
in agreement with previous results.

\end{abstract}

\maketitle


\section{Introduction}
The non equilibrium dynamics of aging and slowly evolving systems
is a topic of current and wide interest \cite{Cugliandolo}.
Much work has been devoted to the understanding of two times
quantities such as the auto-correlation function $C(t,s)=\langle \phi(\vec x,
t)\phi(\vec x,s) \rangle$ and the auto-response function $R(t,s)
=\delta
\langle \phi(\vec x,t) \rangle/\delta h(\vec x,s)$,
where $\phi(\vec x,t)$ is the
order parameter at the space-time point $(\vec x,t)$, $h(\vec x,t)$
is the conjugate external field and the averages are taken over the thermal noise and 
the initial condition with $t \geq s$. The interest in the relation between these two quantities
dates back to the solution by Cugliandolo and Kurchan\cite {Cuglia2} of
the $p$-spin spherical model, where they introduced the fluctuation-dissipation 
relation as a measure of the distance from equilibrium. 
Furthermore, this relation can encode important information on the structure
of the equilibrium state  \cite{Cuglia3}.

One among the simplest examples of systems exhibiting aging and slow dynamics is
a ferromagnetic model evolving with a dissipative dynamics after
a quench from an infinite temperature to a final temperature $T$
smaller than or equal to the critical temperature $T_C$. In both cases, the slow relaxation entails
the separation of the time scales. That is, when
$s$ becomes large enough, the range of $\tau=t-s$ can be devided into the short $\tau \ll s$
and the long $\tau \gg s$ time separation, with quite different behaviors in the
two regimes. The first one is the quasi-equilibrium or stationary regime, where the two time quantities are time 
translation invariant (TTI) and exhibit the same behavior as if equilibrium at the final 
temperature of the quench had been reached. The second one is a genuine off equilibrium regime,
where aging becomes manifest. A crucial point is how these two behaviors are matched.
The generic pattern for phase-ordering systems is that the matching is {\it multiplicative} in the quences to $T_C$
and {\it additive} in the quenches to below $T_C$. The first one is well documented by
analytical calculations. For the second one, although the analytical evidence is less abundant,
the additivity is required on general grounds by the
weak ergodicity breaking scenario \cite{Cugliandolo}.

To be more specific, in the case of the quench to $T_C$, using the methods of the field theoretical
renormalization group (RG), the evolution
equations for $C(t,s)$ and $R(t,s)$ are obtained by means of a
series expansion around the Gaussian fixed point \cite{Schmit,Cala1,Cala1b}. 
The solution of these equations gives for $R(t,s)$  the scaling
form
\be
    R(t,s)=s^{-(1+a)}F_R(t/s)
    \label{scalR}
    \ee
with the additional requirement that the scaling function must be of the form
\be
    F_R(x)=A_R(x-1)^{-(1+a)}x^{\theta}f_R(x) 
    \label{scal2R}
\ee 
where $a=(d-2+\eta)/z$, $d$ is the space dimensionality,
$\eta$ and $z$ are the usual static and dynamic critical exponents, 
$\theta$ is the initial slip exponent and $\lim_{x \to \infty}f_R(x)=1$ \cite{Schmit,Cala1}. 
A similar result is obtained for the correlation function \cite{Schmit,Cala1,Cala1b}.
The multiplicative structure becomes evident rewriting Eq.~(\ref{scalR}) as
\be
     R(t,s)=A_R(t-s)^{-(1+a)}g_R(x) 
    \label{scal3R}
\ee 
where $g_R(x)=x^{\theta}f_R(x)$.

In the case of the quench to $T< T_C$, the above form is replaced by the additive structure
\be
     R(t,s)=R_{st}(t-s) + R_{ag}(t,s)
    \label{splitR}
\ee 
where the first is the stationary contribution and the second one is the aging contribution
which obeys a scaling form of the type~(\ref{scalR})
\be
     R_{ag}(t,s)=s^{-(1+a)}h_R(x)
    \label{scal4R}
\ee 
without the restriction~(\ref{scal2R}) on the form of $h_R(x)$.

In the latter case, all theoretical efforts have been direct toward the determination of the exponent $a$ and
the scaling function $h_R(x)$. Keeping into account that the evolution is 
controlled by the $T=0$ fixed point, which in no case is Gaussian, the perturbative RG cannot be used
and resorting to uncontrolled approximations is unavoidable.
Among these, one of the most successful is the Gaussian auxiliary field
(GAF) approximation. The method was originally introduced by Ohta,
Jasnow and Kawasaki \cite{OJK} in the study of the scaling behavior of
the structure factor and has been subsequently applied also to the study of the response function
\cite{Berthier,noiachi}.  Recently, new
results for $R(t,s)$ have been obtained by Mazenko \cite{Mazenko} using 
a perturbative expansion which improves on the GAF approximation.
Next to approximate methods, there exist exact analytical
results for two solvable models: the one dimensional Ising
model \cite{Lippiello,God2} and the ${\cal O}(N)$ model in the large $N$ limit for
arbitrary dimensionality \cite{noininfi}. 
Both solutions give for $R_{ag}(t,s)$ the scaling form (\ref{scal4R}), with
$a=0$ for the $d=1$ Ising model and $a=(d-2)/2$ for the  large $N$
model.

Therefore, in the context of the quenches to below $T_C$, where a controlled theory is not available
and numerical simulations are very  time
demanding, it is of much interest the conjecture put forward by Henkel et al. \cite{LSI,LSI2}
that the  response function transforms covariantly under the group of
local scale transformations, {\it both} in the quenches to and to below $T_C$. The hypothesis of
local scale invariance (LSI), then, implies that the multiplicative structure for $R(t,s)$, as obtained from RG
arguments at $T_C$, applies also in the quenches to below $T_C$. That is, 
from LSI follows that $R(t,s)$ obeys Eq.~(\ref{scal3R}), 
both at and below $T_C$, with the additional prediction that
\be
  f_R(x)\equiv 1 
  \label{lsi}
  \ee
holds not just asymptotically, but for all values of $x$,
while the amplitude $A_R$ and  the exponents $a$ and $\theta$ remain unspecified.
Hence, whith the LSI hypothesis, the difference between the quenches
to $T_C$ and to below
$T_C$ would be left only in the values of the exponents $a$ and $\theta$. This
is actually verified by the exact solution of the
spherical model \cite{noininfi,God}.  Conversely, from the GAF
approximation and from the exact solution of the $d=1$ Ising model 
follows \cite{noiTRM}  
\be
f_R(x)=[(x-1)/x]^{1/z}
\label{100.1}
\ee
which differs significantly from the above LSI prediction~(\ref{lsi}).

In the case of the quench to $T=T_C$, the validity of LSI 
has been tested by Calabrese and Gambassi \cite{Cala2} by means of the 
$\epsilon$ expansion.
Their field theoretical
computation shows that LSI holds up to the first order in 
$\epsilon =4-d$, but deviations of order $\epsilon ^2$ are
present. Motivated by this result, Pleimling and Gambassi (PG) in a
recent paper \cite{Pleimling} have carried out a careful numerical check of
both LSI based and field theoretical calculations in the Ising model
quenched to $T_C$, in $d=2$ and $d=3$. In particular, they have computed
the integrated global response to a uniform external field, finding i)
a discrepancy between the LSI behavior and the data, ii) that the discrepancy is
more severe in $d=2$ than in $d=3$ and iii) that the $\epsilon^2$ correction 
does not eliminate the discrepancy, but improves
on the LSI prediction. In this connection, Calabrese and Gambassi \cite{Cala1b} 
first and then PG made the remark that
the LSI prediction coincides with the Gaussian approximation, thus accounting
for the agreement between LSI and the solution of the spherical model.

Following through this suggestion, 
one could anticipate that the discrepancy
between the LSI behavior and the results of simulations should disappear in the quench 
to $T_C$ with $d=4$, while it ought to get even worse in the quench to below $T_C$, independently
of the dimensionality. Furthermore, in the latter case the failure of LSI is expected to
be not just in the quantitative accuracy of the approximation, but also of a structural
character since the multiplicative form of $R(t,s)$ is incompatible with the weak ergodicity
breaking scenario.

In order to investigate these ideas, one can take advantage of the
efficient numerical tools made available by a new generation of
algorithms \cite{Chat1,Ricci,noialg}. 
These algorithms are based on the relation between $R(t,s)$ and unperturbed
quantities which, by speeding up the simulation, allow for the
measurement of $R(t,s)$. In this paper,
exploiting the algorithm  introduced by us \cite{noialg}, we extend the
investigation of the Ising model carried out by PG
to the two cases of the quenches to
$T_C$ with $d=4$ and to below $T_C$ with $d=2$. Rather than
computing the integrated response function for a global quantity, as PG have done, we access directly the
local response function $R(t,s)$, thus making the
comparison between the numerical data and Eq.(\ref{lsi}).  

In the $d=4$ Ising model quenched to
$T=T_C$, after addressing the question of the universality  
of the exponent $\theta$ \cite{theta} and of the ratio $T_C A_R/A_C$
between the amplitudes of response and correlation function \cite{Cala1,Cala1b,God,Chat}, 
we find an excellent quantitative agreement between
the numerical data and the analytical results from the Gaussian model. 
In particular, we find that
both for $R(t,s)$ and $C(t,s)$ 
not only the scaling exponents, but also the scaling functions and the
ratio  $T_CA_R/A_C$  are well accounted for in  the Gaussian approximation. 
We find that Eq.~(\ref{lsi}) holds
and we conclude that LSI correctly describes the critical quench of the $d=4$ Ising
model. 
Conversely, in the quench of the  $d=2$ Ising model
to $T=T_C$ and to $T<T_C$, important
deviations from LSI are observed. These findings do bring 
support to the idea that the LSI principle is some sort of zero
order theory of Gaussian nature and contradict previous statements \cite{Abriet,Pleimling}
that no deviations from LSI predictions are observed in the measurements of
local quantites.

The paper is organized as follows. In sec.\ref{sec1} we shortly review existing
results for $C(t,s)$ and $R(t,s)$. In particular in sec.\ref{sec1A}
and in sec.\ref{sec1B} we give 
the results from RG arguments and from the Gaussian model, respectively, while 
in sec.\ref{sec1C} we present a phenomenological picture for the quench to $T<T_C$. In
sec.\ref{sec2} we outline the algorithm used in the
simulations and in sec.\ref{sec3} we present and discuss the numerical results. 
Concluding remarks are made in the last section.

\section{Existing results}
\label{sec1}

We consider a system with a non conserved scalar order parameter $\phi(\vec x,t)$ 
(model A in the classification of Hohenberg and Halperin~\cite{Hohe}) 
evolving with the Langevin equation
\be
\frac{\partial \phi(\vec x,t)}{\partial t}= -\frac{\delta
  H[\phi]}{\delta \phi(\vec x,t)}+\eta(\vec x,t)
\label{langevin}
\ee
where $\eta(\vec x,t)$ is a Gaussian white noise with expectations 
\be
\langle \eta(\vec x,t)\rangle= 0 \hskip 1cm
\langle \eta(\vec x,t)\eta(\vec x',t')\rangle =2T
\delta(\vec x-\vec x')\delta(t-t')
\ee
 and $H[\phi]$ is of 
the Ginzburg-Landau-Wilson form 
\be
H[\phi]=\int d \vec x \left [\frac{1}{2} (\vec \nabla \phi)^2+
\frac{1}{2} r \phi^2+\frac{1}{4!} g \phi^4 \right ] ,
\label{gaussian}
\ee
with $r<0$ and $g > 0$.
The system is prepared in an uncorrelated Gaussian initial state with
expectations 
\be
\langle \phi(\vec x,0)\rangle= 0 \hskip 1cm
\langle \phi(\vec x,0)\phi(\vec x',0)\rangle =\tau_0^{-1}
\delta(\vec x-\vec x').
\ee

\subsection{Quench to $T_C$: RG results}
\label{sec1A}

In the case of the quench to $T_C$
one can show, by means of standard RG methods \cite{Schmit,Cala1,Cala1b},
that $\tau_0^{-1}$ is an irrelevant variable. Thus, putting
$\tau_0^{-1}=0$, one obtains the leading scaling behavior which is given by 
Eqs.~(\ref{scalR},\ref{scal2R}) for $R(t,s)$, whereas for the
correlation function one has   
\be
    C(t,s)=s^{-b}F_C(t/s) ,
    \label{scalC}
    \ee
with the scaling function 
 \be
    F_C(x)=A_C(x-1)^{-b}x^{\theta -1}f_C(x),
    \label{scal2C}
\ee and
 \be
 b=a=\frac{d-2+\eta}{z} .
    \label{a1}
    \ee
As for $f_R(x)$, the RG method allows to fix only 
the large $x$ behavior
$\lim_{x \to \infty}f_C(x)=1$.

From Eq.(\ref{a1}) one has that $a$ and $b$ are related to the critical
exponents $\eta$ and $z$. Therefore,
according to the classification of Hohenberg and Halperin~\cite{Hohe}, 
$a$ and $b$ take the same value for systems belonging
to the same class of universality. 
The problem  of the univerality of $\theta$ has been addressed in a
series of papers \cite{theta,Chat}. Furthermore,  
Godr\`eche 
and Luck \cite{God} have proposed that also the ratio 
$T_C A_R/A_C$ is a universal quantity.
More precisely,  considering the limit fluctuation dissipation ratio \cite{Cuglia2}  
$X_\infty$ defined by
\be
   X_\infty=\lim_{s\to \infty}\lim_{t\to \infty} \frac{T R(t,s)}
    {\partial _sC(t,s)} 
\label{xinfty0}
   \ee
and using Eqs.(\ref{scalR},\ref{scal2R},\ref{scalC}), in the quench to the critical point one has 
\be
   X_\infty=\frac{T_CA_R}{A_C(1-\theta)}.
\label{xinfty}
   \ee
Universality of  $\theta$ and $T_C A_R/A_C$
implies  universality of $X_\infty$.
Indeed, we will see that numerical results
for $\theta$ and $X_\infty$, in the $d=4$ Ising model,
give the same values as in the Gaussian model.

\subsection{Quench to $T_C$: the Gaussian model}\label{sec1B}

The critical Gaussian model is obtained putting $r=0$ and $g=0$ in the Hamiltonian~(\ref{gaussian}).
Then, the equation of motion (\ref{langevin}) can be
solved in Fourier space yielding 
\be
    C(\vec k,t,s)=\langle\phi(\vec k,t)\phi(-\vec k,s)\rangle =
    \frac{T_C}{k^2}\left [e^{-k^2(t-s)}-e^{-k^2(t+s)}\right ]
    +\frac{e^{-k^2(t+s)}}{\tau_0}
\label{fou1}
\ee
\be
    R(\vec k,t,s)=\frac{\delta\langle\phi(\vec k,t)\rangle}{\delta
      h(\vec -k,s)} =e^{-k^2(t-s)} 
\label{fou2}
\ee
whith $t>s$.
The auto-correlation function and the auto-response function are 
obtained integrating over $\vec k$ the above equations. In order to
regularize the equal time behavior of $C(t,s)$ and $R(t,s)$ one must
introduce an high momentum cut-off and, for simplicity, we choose
a smooth cut-off implemented by the multiplicative factor $e^{-k^2/\Lambda^2}$ in
Eqs.(\ref{fou1},\ref{fou2}). 
Neglecting the last term in Eq.(\ref{fou1}), in order to keep
only the leading scaling behavior, one gets \cite{CugliaParisi}  
\be
C(t,s)=\frac{2 T_C}{(d-2)(4\pi)^{d/2}}
\left [(t-s+t_0)^{1-d/2}-(t+s+t_0)^{1-d/2}\right ],
\ee
\be
R(t,s)=\frac{1}{(4\pi)^{d/2}}
(t-s+t_0)^{-d/2}
\ee
where $t_0=1/\Lambda^2$.
Notice that the specific choice of the cut-off affects
the behavior of $R(t,s)$ and $C(t,s)$ only on the time scale $t-s \simeq
t_0$. Taking $t-s \gg t_0$,    
the above results are in the scaling form of
Eqs.~(\ref{scal2C},\ref{scal2R}) with $f_R(x)\equiv 1$, as required by LSI, and with
$a=b=d/2-1$, $\theta =0$, $f_C(x)=x- x(x-1)^{a}(x+1)^{-a}$. 
In particular, in $d=4$ one has 
\be
C(t,s)=A_C s^{-1}(x-1+t_0/s)^{-1}(x+1+t_0/s)^{-1} \label{d41} \ee
\be
R(t,s)=A_R(t-s+t_0)^{-2}  , \label{d43} \ee 
with $A_C=2T_C/(4\pi)^2$ and $2T_C A_R=A_C$.

\subsection{Quench to below $T_C$: phenomenological picture}
\label{sec1C}

In the case of the quench to below $T_C$, the system evolution is
characterized by the formation and subsequent growth of compact
ordered domains whose typical size increases with the power law
\be
L(t) \sim t^{1/z}.
\label{growth}
\ee
The evolution via domain coarsening suggests
\cite{Cugliandolo}, for large $s$, the additive form of
the correlation function
\be
    C(t,s)=C_{st}(t-s)+C_{ag}(t,s)
    \label{splitC}
    \ee
where $C_{st}(t-s)$  represents the correlation function of the equilibrium
fluctuations within an infinite domain and $C_{ag}(t,s)$ is the domain walls contribution. 
Analytical solutions \cite{noininfi,God}
as well as numerical results \cite{noiRd2}  confirm this
structure, with $C_{ag}(t,s)$ obeying a scaling form as in
Eq.(\ref{scalC}) and with $b=0$.
As stated in the Introduction, the similar structure~(\ref{splitR}) holds also 
for the response function
with $R_{ag}(t,s)$ in the scaling form~(\ref{scal4R}) and 
$R_{st}(t-s)$ related to $C_{st}(t-s)$ by the fluctuation dissipation theorem,
$T R_{st}(t-s)= \partial C_{st}(t-s)/\partial s$.

Numerical simulations \cite{noiachi,noiTRM} for the zero field cooled magnetization
$\chi(t,t_w)=\int_{t_w}^{t}R(t,s)ds$  
are consistent with the additive
structure~(\ref{splitR}) 
and with a scaling function in Eq.(\ref{scal4R}) of the form
\be
    h_R(x)=A_R\frac{x^\beta}{(x-1+t_0/s)^{1-1/z+a}}
    \label{GAF}
    \ee
where $A_R,a,\beta,t_0$ are phenomenological parameters, while $z$ is the dynamical exponent entering 
Eq.~(\ref{growth}). Here, $t_0$ is a microscopic time which is negligible except when $x \rightarrow 1$.
Recent results \cite{noiRd2} from the direct measurement 
of $R(t,s)$ do support the above form of $h_R(x)$  and give a 
quantitative estimate of $a$, $A_R$ and $\beta$.  
The physical meaning of Eq.~(\ref{GAF}) becomes clear
for short time separation $t-s \ll s$. In this case
Eqs.~(\ref{scal4R},\ref{GAF}) can be rewritten as
\be
R_{ag}(t,s)=\rho_I(s) R_{sing}(t-s)
\label{GAF.1}
\ee
where
\be
R_{sing}(t-s)=A_R(t-s+t_0)^{-1-a+1/z}
\ee
and $\rho_I(s) \propto L^{-1}(s)$ is the interface density at time $s$.
Therefore, $R_{sing}(t-s)$ represents the response of a single interface and
Eq.~(\ref{GAF.1}) simply means that the aging contribution in the
response is produced by the interfacial degrees of freedom.
For larger time separation $t-s >s$, interfaces interact with each other
and the interaction generates the term $x^\beta$ in Eq.~(\ref{GAF}).
The form (\ref{GAF}) of $h_R(x)$ is corroborated by  the exact analytical result for the
$d=1$  Ising model with non conserved order parameter 
\cite{Lippiello,God2}, by the
numerical results of the $d=1$ Ising model with
conserved order parameter \cite{noialg} and 
by the analytical results obtained with
the GAF approximation \cite{Berthier,noiachi}.
In the sec.\ref{sec3D} we will present a direct comparison between 
Eq.~(\ref{GAF}) and the prediction from LSI.

\section{The algorithm}\label{sec2}

We consider a system of $N$ spins on a lattice with the Ising Hamiltonian
\be
H=-J \sum_{<ij>}\sigma_i \sigma_j
\ee
where the sum runs over the nearest neighbours pairs $<ij>$ and $J>0$. 
The time evolution is then obtained through single spin flip dynamics with Glauber
transition rates
\be
w_i([\sigma] \to [\sigma'])=\frac{ 1}{2}
\left[ 1-\sigma_i \tanh\left(\frac{h_i^W}{T}\right)\right]
\label{w}
\ee
where $[\sigma]$ and $[\sigma']$ are spin configurations differing
only for the value of the spin in the $i$-th site,
$h_i^W=J\sum_k^{\prime}\sigma_k$ is the Weiss field, $J$ is the ferromagnetic coupling and the sum is restricted to
the nearest neighbors of the $i$-th site. 
$C(t,s)$ and $R(t,s)$ are given by
\be
C(t,s)=\frac{1}{N}\sum_i\langle \sigma_i(t)\sigma_i(s)\rangle
\ee
and
\be
R(t,s)=\lim_{\Delta s \to 0} \frac{1}{\Delta s N} \sum_i
\left . \frac{\partial \langle \sigma_i(t)\rangle }{\partial h_i} \right
\vert_{h=0} 
\label{rts}
\ee
where $\sigma_i(t)$ is the spin in the $i$-th site at time t and
$h_i$ is an external field acting on the $i$-th site during 
the time interval $[s,s+\Delta s]$.
In the computation of $R(t,s)$, we use our own algorithm \cite{noialg}, 
which offers higher efficiency with respect 
to other methods \cite{Chat1,Ricci} allowing to compute the response function
without imposing the external field. Carring out 
the derivative in Eq.~(\ref{rts}) we find
\be
T R(t,s)=\frac{1}{2}\lim _{\Delta s \to 0}\left [
\frac {C(t,s+\Delta s)-C(t,s)}{\Delta s}-\langle \sigma _i(t-\Delta s)B_i(s)\rangle \right ]
\label{treb}
\ee
where $B_i$ enters the evolution of the magnetization according to \cite{noialg}
\be
\frac {d\langle \sigma _i(t)\rangle}{dt}=\langle B_i(t)\rangle .
\ee  
The above result is quite general and is independent of the
details of the Hamiltonian and of the transition rates. Furthermore, it
can be easily generalized to the case of vector order parameter 
\cite{noiclock}. 
In the case of single spin flip dynamics, one has 
\be 
B_i(t)=2\sigma _i(t)w_i([\sigma ]\to [\sigma _i])
\ee
with $w_i([\sigma ]\to [\sigma _i])$ given in Eq.(\ref{w}).

In order to improve the signal to noise ratio, we compute 
the quantity
\be
\mu(t,[s+1,s])=\int _{s}^{s+1} R(t,t')dt'
\label{irff}
\ee
which is the response to a perturbation acting in the time window $[s,s +1]$. 
Here and in the following we express time in
units of a Monte Carlo step. Replacing the integral in Eq.~(\ref{irff})
by a discrete summation
on the microscopic time scale $\epsilon=1/N$, from Eq.~(\ref{treb})
we obtain
\be
T\mu (t,[s+1,s])=
\frac{1}{2}\left [C(t,s+1)
-C(t,s)-\frac{1}{N}\sum_{i, k=1}^N
\langle \sigma _i(t-1/N)B_i(s+(k-1)/N)\rangle \right ] .
\label{trec}
\ee 
Because of the scaling form (\ref{scalR}) for $R(t,s)$, it is easy to
show \cite{noiRd2} that $R(t,s)$ coincides with $\mu (t,[s+1,s])$ up to
corrections of order $1/s$ which, in the considered range of times, can
always be  neglected. Therefore, in the simulations we 
identify $R(t,s)$ with $\mu (t,[s+1,s])$ and the numerical results for
$R(t,s)$ are obtained from Eq.~(\ref{trec}). In all cases we take 
a completely disordered initial state which, in principle, produces a correction
to scaling. However, this is not detectable in the
time region explored in the simulation.

\section{Numerical results}\label{sec3}

\subsection{$d=4$,   $T=T_C$}

We have considered a system of $N=60^4$ Ising spins on a four
dimensional hypercubic
lattice  
quenched  to the critical
temperature $T_C \simeq 6.68 J$ \cite{tcd4}. The response
and correlation functions are then computed for four different values of 
$s=25,50,75,100$. In all Figures the error bars are smaller than the symbols.

\begin{figure}
    \centering
\includegraphics[width=12cm]{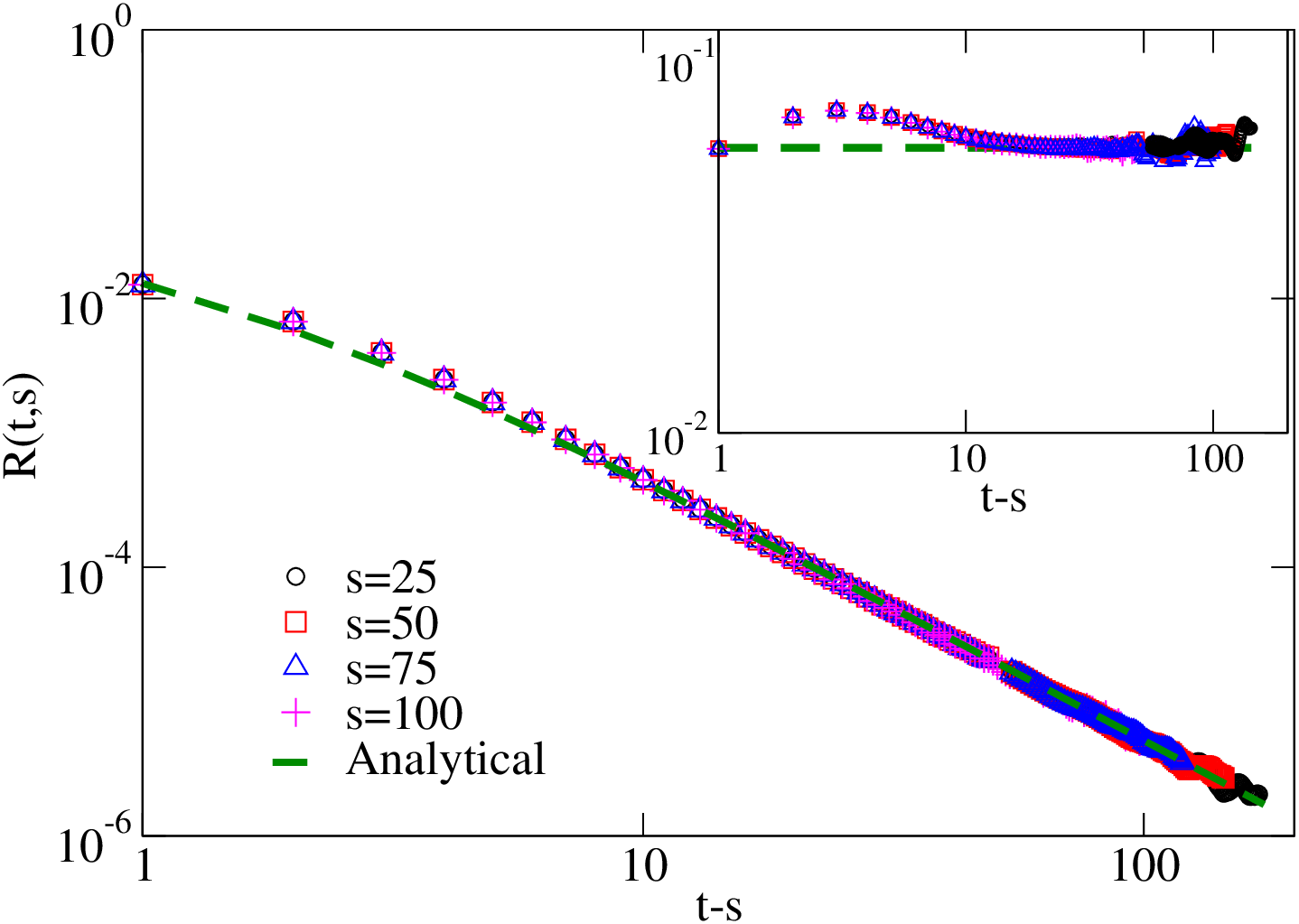}    
    \caption{(Color on line)
  $R(t,s)$ in the $d=4$ Ising model
      quenched to $T=T_C\simeq6.68J$. The broken line is the
     analytical solution of the Gaussian model (Eq.(\ref{d43})) with
        $A_R=0.35/T_C$ and $t_0=0.1$. \\
Inset: plot of $(t-s+t_0)^2R(t,s)$ showing
     the absence of corrections to Gaussian behavior. The broken line indicates the constant value.}
    \label{fig41}
\vskip2cm
\end{figure}

In order to compare with  the results of the
Gaussian model given in~\ref{sec1B}, we observe that $R(t,s)$ in Eq.~(\ref{d43})
depends only on the time difference $t-s$.
This result is
reproduced in the numerical simulations. Indeed, plotting the
curves for different $s$ as a function of $t-s$ (see
Fig.\ref{fig41} we find the collapse on a master curve that
is well described by Eq.~(\ref{d43}), with $T_C A_R=0.35 \pm 
0.02$ and $t_0\simeq0.1$. There is a very small difference
only for short
time separations $t-s \sim 1$, which can be attributed to the specific
choice of a smooth cut-off used in the integration over
$\vec k$ of Eq.~(\ref{fou2}). 

\begin{figure}
    \centering
\includegraphics[width=12cm]{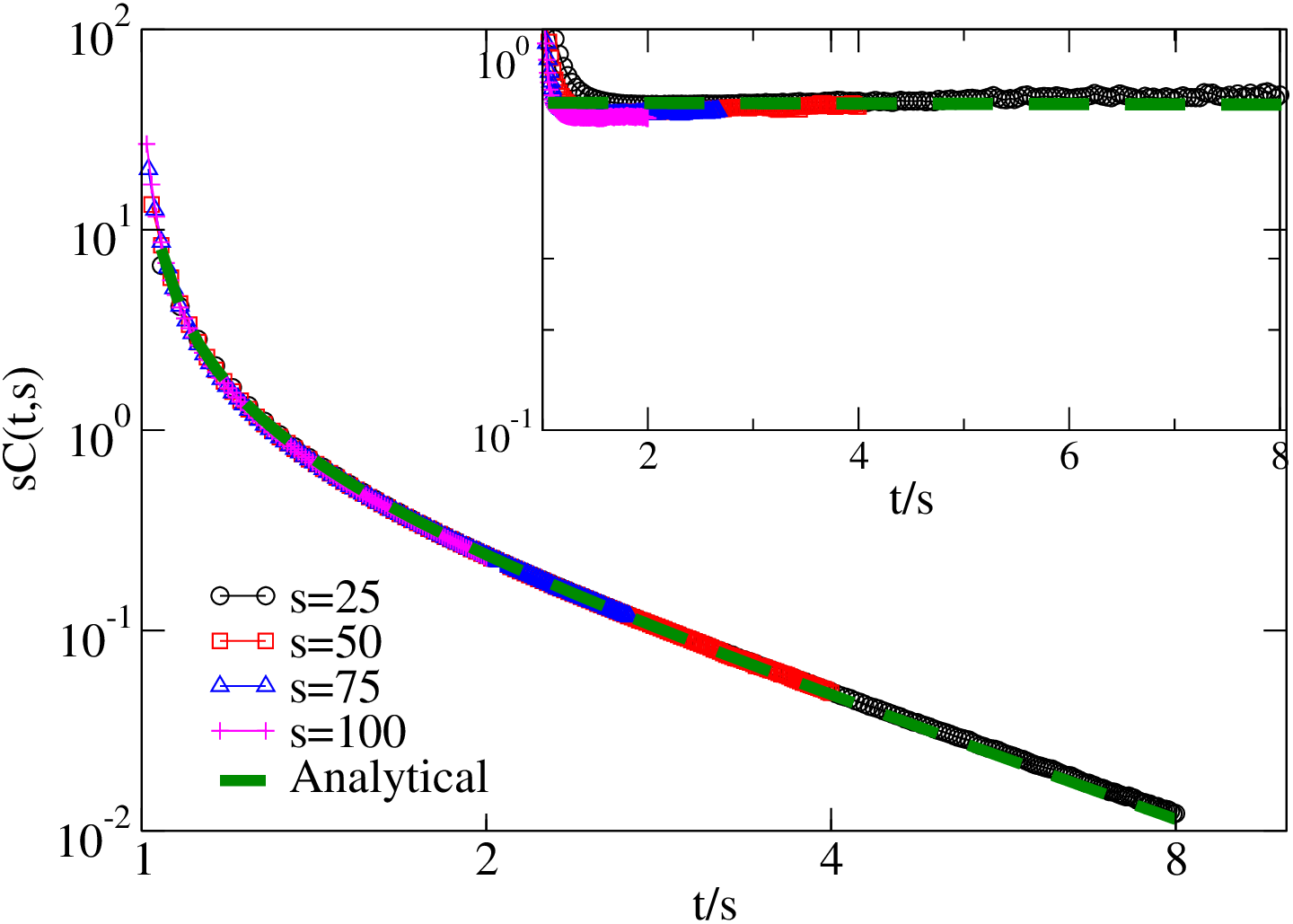}    
    \caption{(Color on line) $C(t,s)$ in the $d=4$ Ising model
      quenched to $T=T_C\simeq6.68J$. The broken line is the
     analytical solution of the Gaussian model (Eq.(\ref{d41}) with
        $A_C=0.72$ and $t_0=0.1$. \\
Inset: plot of $(x-1+t_0/s)(x+1+t_0/s)sC(t,s)$ showing
     the absence of corrections to Gaussian behavior. The broken line indicates the constant value.}
\label{fig42}
\end{figure}

In Fig.\ref{fig42} we compare the
numerical data for $C(t,s)$ with the analytical expression of 
Eq.~(\ref{d41}). The plot shows that the quantity $sC(t,s)$ 
depends only  on the ratio $t/s$ in complete agreement with
Eq.~(\ref{d41}), where $t_0/s$ can be neglected and with $A_C=0.72 \pm 0.02$. 

Both Fig.\ref{fig41} and Fig.\ref{fig42} 
show the accuracy of the numerical
method and that the Gaussian approximation gives the
correct results for $C(t,s)$ and $R(t,s)$ in $d=4$ and
$T=T_C$. Furthermore, according to universality, the numerical values for $T_C$, $A_R$ and $A_C$ yield
an amplitude ratio in agreement,
within numerical uncertainty,  with the Gaussian result $T_CA_R/A_C=1/2$.
No logarithmic corrections are numerically detected in the range of times
investigated, as shown in the insets of Fig.\ref{fig41} and Fig.\ref{fig42}.

\subsection{$d=2$, $T=T_C$}\label{sec3B}

\begin{figure}
    \centering
\includegraphics[width=12cm]{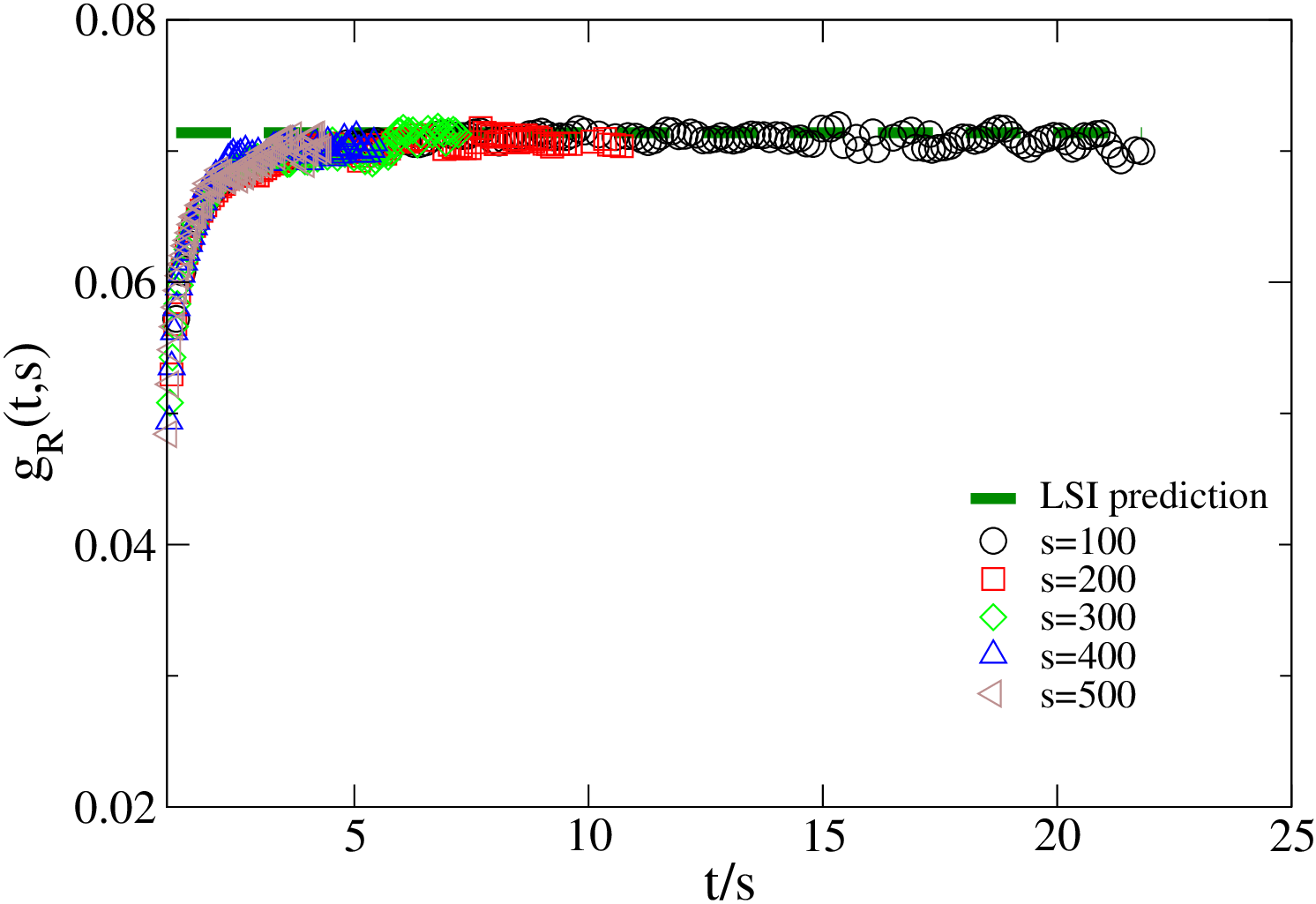}    
    \caption{(Color on line) $g_R
(t,s)$ defined in Eq.(\ref{gr}) with
    $\theta=0.38$ and $a=0.115$ in the
    the $d=2$ Ising model
      quenched to $T=T_C\simeq 2.26918J$. The broken line is the prediction
      of LSI.}
\label{fig21}
\vskip2cm
\end{figure}

\begin{figure}
    \centering
\includegraphics[width=12cm]{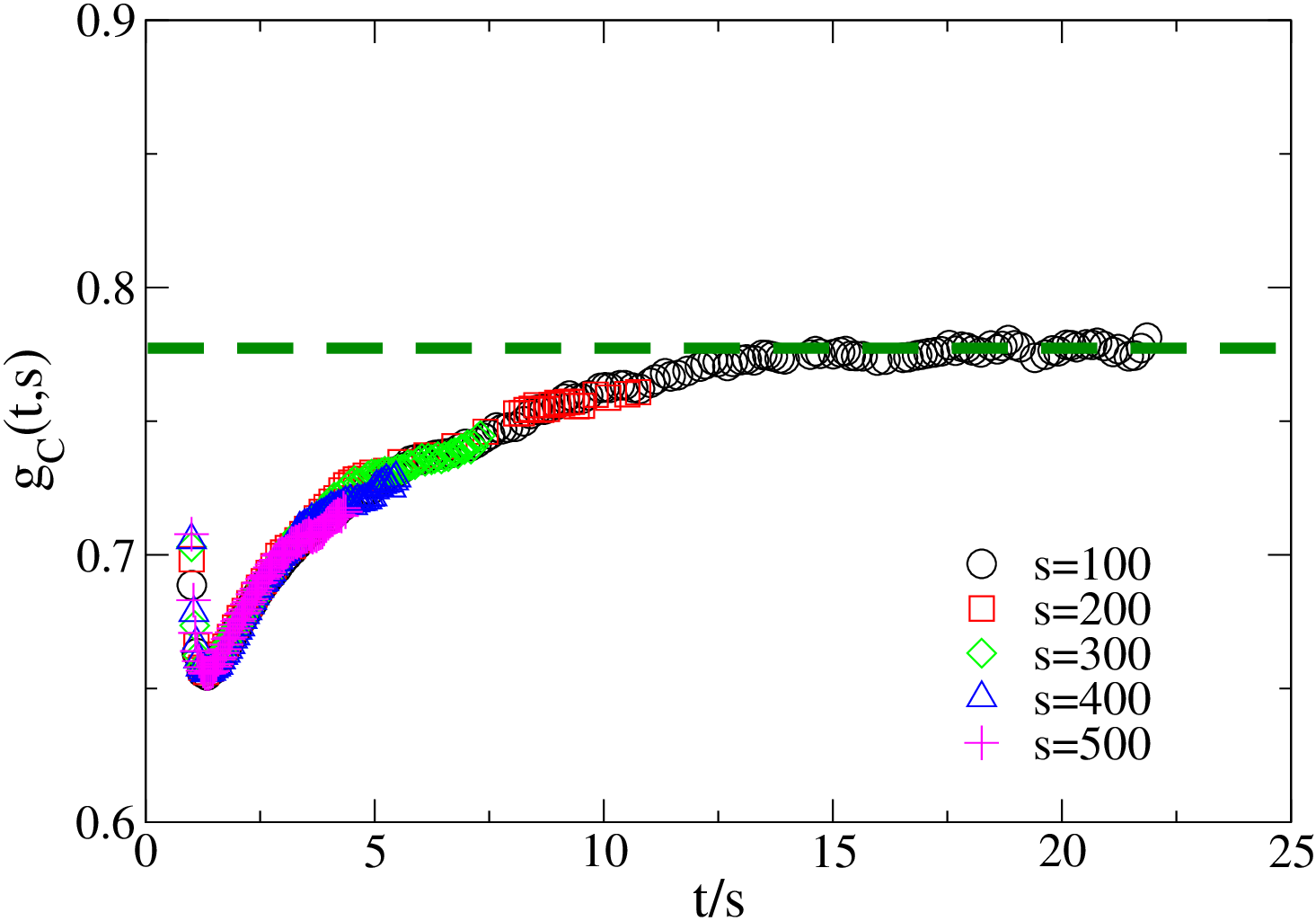}    
    \caption{(Color on line) $g_C(t,s)$ defined in Eq.(\ref{gc}) with
    $\theta=0.38$ and $a=0.115$ in the
    the $d=2$ Ising model
      quenched to $T=T_C\simeq 2.26918J$.  The broken line corresponds
    to the amplitude $A_C =0.78$. }
\label{fig22}
\vskip2cm
\end{figure}

We consider a square lattice with $N=1000^2$ Ising spins and we compute
numerically the response and correlation functions in the quench to
$T_C\simeq 2.26918J$, for five different values of 
$s=100,200,300,400,500$. In Fig.\ref{fig21} and Fig.\ref{fig22} we have plotted the quantities
\be
g_R(t,s)=(t-s)^{a+1}(t/s)^{-\theta}R(t,s)
\label{gr}
\ee
\be
g_C(t,s)=(t-s)^b(t/s)^{1-\theta}C(t,s)
\label{gc}
\ee 
versus $t/s$, with  $\theta=0.38$ and $a=b=0.115$ taken from ref. \cite{Pleimling}. 
We find data collapse as expected from the RG results of  Eqs.~(\ref{scal2R},\ref{scal2C})
which yield
\be
g_C(t,s)=A_Cf_C(t/s), \hspace{0.3cm}  g_R(t,s)=A_Rf_R(t/s). 
\ee 
Using, next, the asymptotic conditions $\lim_{x \to \infty}f_R(x)= \lim_{x \to \infty}f_C(x) = 1$
we can extract the amplitudes $A_C=0.78 \pm 0.02$ and
$A_R= 0.071 \pm 0.002 $. Lastly, from Fig.\ref{fig21} we can make a
check on the LSI prediction (\ref{lsi}) that $g_R(t,s)$ ought to be constant with
$g_R(t,s) \equiv A_R$.
Fig.\ref{fig21} shows that there is an evident deviation from LSI 
for $x<5$, while the LSI behavior holds for $x>5$. 

\subsection{The limit fluctuation dissipation ratio $X_\infty$}

We measure $X_\infty$ using Eq.~(\ref{xinfty}) with values for $A_C$, $T_C A_R$ and $\theta$ estimated
from the numerical data.
In $d=4$, with  $\theta=0$, $A_C=0.72\pm0.02$ and $T_C A_R=0.35\pm
0.02$, we find $X_\infty=0.49 \pm 0.03$ in agreement 
with the Gaussian result $X_\infty=1/2$.
 This supports the idea that
not only  $\theta$, but also $X_\infty$ is universal \cite{God,Cala1,Cala1b}.

In $d=2$, we find $A_C=0.78\pm0.02$, $T_C A_R=0.161 \pm 0.001$ and taking $\theta=0.38$ from Ref.~\cite{Pleimling}
we obtain $X_\infty=0.33\pm 0.01$, in
agreement with previous numerical results obtained with different methods
\cite{Chat,Sollich} and with $X_\infty=0.30\pm 0.05$ from the two loop $\epsilon$
expansion \cite{Cala2}. For convenience, the numerical values of exponents, amplitude ratio and 
$X_\infty$ in the different processes have been collected in Table \ref{tabella}.

\subsection{$d=2$, $T<T_C$}\label{sec3D}

In the quench to below $T_C$ the behavior of the data in the short time separation regime $t-s \ll s$
allows to discriminate between the additive and the multiplicative forms of $R(t,s)$.
Expanding up to first order in $(t-s)/s$, in the former case from from Eqs.~(\ref{splitR}) and~(\ref{scal4R})
one ontains
\be
R(t,s) = R_{st}(t-s) + s^{-(1+a)}\left[ h_R(1) +h^{\prime}_R(1) \left ({t-s \over s}\right) \right]
\label{add.1}
\ee
while from the LSI form  
\be
R_{LSI}(t,s)=A_R(t-s)^{-(1+a)}x^\theta
\label{ins.1}
\ee
one gets
\be
R_{LSI}(t,s)=A_R(t-s)^{-(1+a)}\left[ 1+ \theta \left ({t-s \over s} \right) \right].
\label{ins.2}
\ee
Therefore, as $t-s$ becomes small with finite $s$, from Eq.~(\ref{add.1}) there remains
an $s$ dependence due to $s^{-(1+a)}h_R(1)$, while from Eq.~(\ref{ins.2}) there is no residual
$s$ dependence.

\begin{figure}
    \centering
\includegraphics[width=12cm]{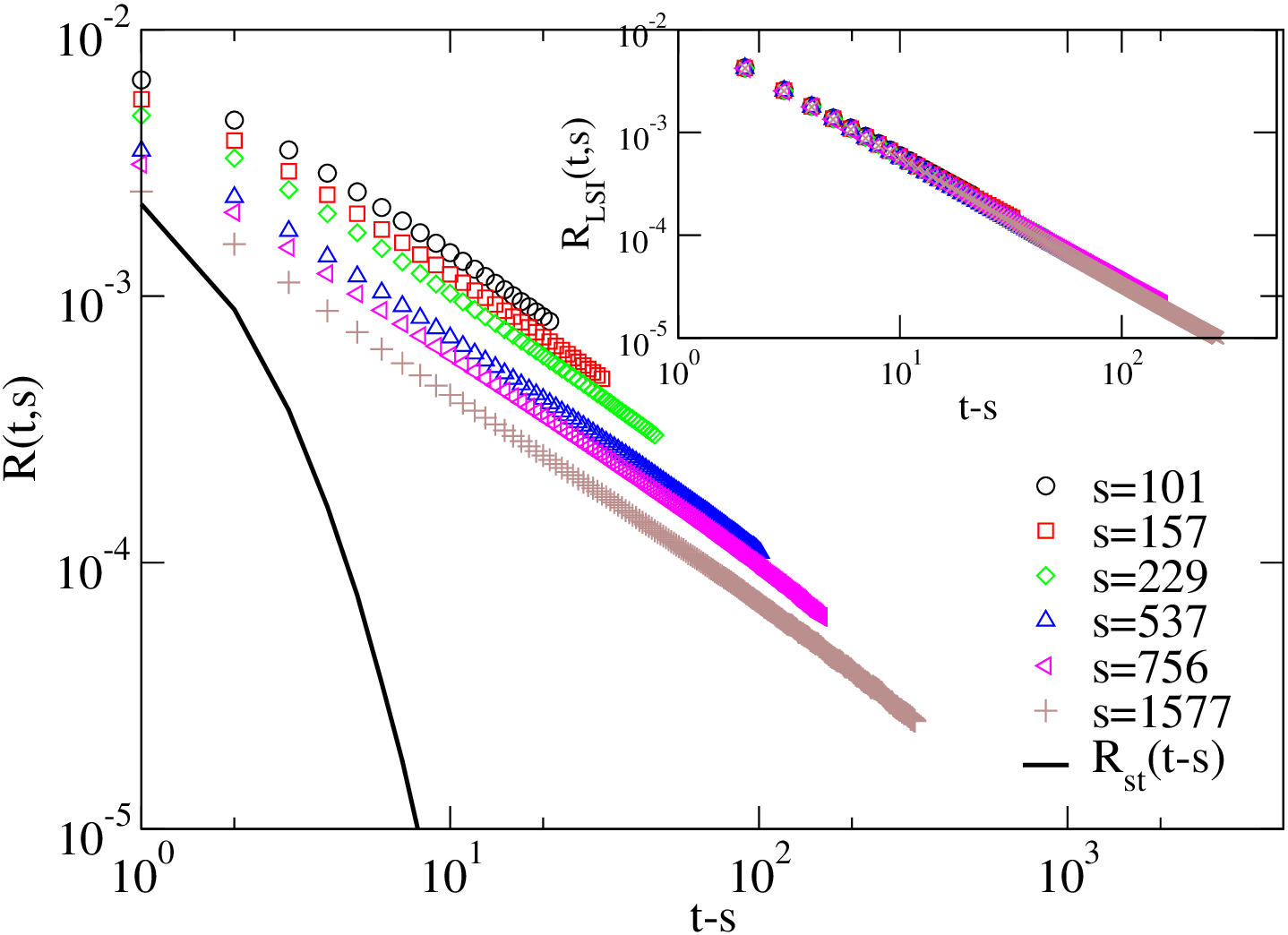}
    \caption{(Color on line) $R(t,s)$ in the quench of the Ising model
      to $T=1.5J$  below $T_C$. Inset: plot of $R_{LSI}(t,s)$ for the same values
of $s$ and $t-s$. The continous line is the plot of $R_{st}(t-s)$.}
\vskip2cm
\label{fig23}
\end{figure}

In order to see which of the two behaviors is actually realized in the data, we have quenched
a system of $1000^2$ Ising spins to the temperature
$T=1.5J \simeq 0.69T_C$, 
taking the wide range of $s \in [101,1577]$ and
focusing on the regime $t-s \le s$. The numerical data are displayed in
Fig.\ref{fig23}. Furthermore, in the inset we have plotted Eq.~(\ref{ins.1})
in the same range of $s$ and $t-s$, with $A_R=0.01$ obtained by imposing 
$R(s+1,s)=R_{LSI}(s+1,s)$ for $s=100$, $a=0.27$ extracted from the data for
$R(t,s)$ (see below) and $\theta=\lambda/z-1-a$ with $\lambda/z=0.625$ \cite{LSI}.
The numerical data display an evident dependence on $s$ down to $t-s=1$, which is absent in those
for $R_{LSI}(t,s)$ (note the same vertical scale). Therefore, the LSI form of $R(t,s)$
can be ruled out.

We have also extracted $R_{st}(t-s)$ from the data using
the following protocol. We have let the system to evolve in contact with the
thermal reservoir at the
temperature $T=1.5J$ after preparing it in a completely ordered
state, for instance all spins up. The equilibration time $t_{eq}$ for this process is finite 
and $R_{st}(t-s)$ is obtained by measuring the response function for $s>t_{eq}$.
The data obtained in this way yield, as expected, an exponentially decaying contribution
(continous line in Fig.\ref{fig23}),
which becomes very rapidly negligible with respect to
the full $R(t,s)$. Therefore, in the observed range of $s$ and $t-s$,
i) aging is well developed in the data and is due to the $R_{ag}(t,s)$ contribution
in Eq.~(\ref{splitR}), while it is practically
unobservable in the LSI and ii) the stationary contribution from the data decays
exponentially, while in the LSI there is a power law decay.

\begin{figure}
    \centering
     \includegraphics[width=12cm]{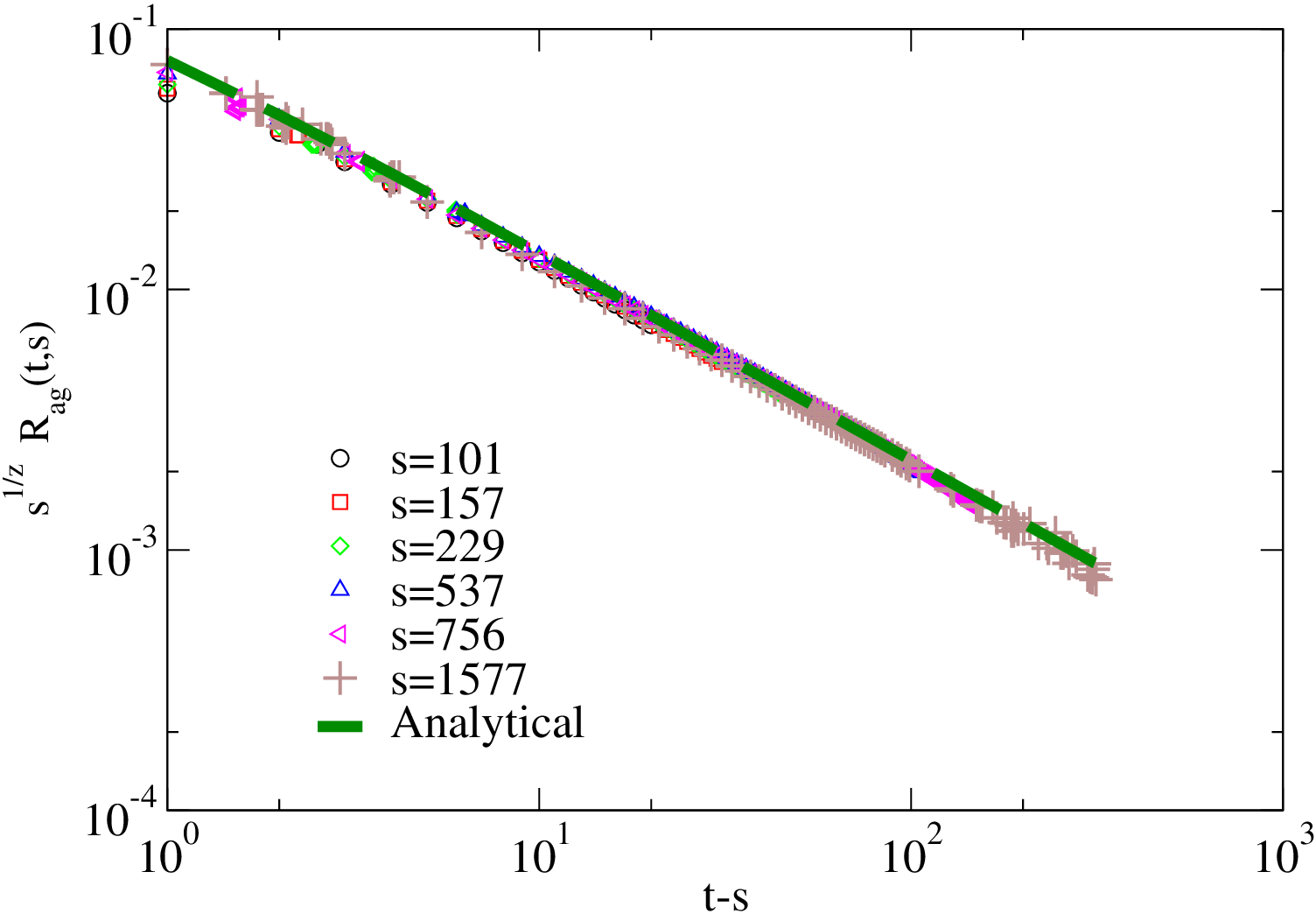}
    \caption{(Color on line)  $s^{1/z} R_{ag}(t,s)$ with $z=0.47$ for the
      $d=2$ Ising model quenched to $T=1.5J$ below $T_C$. The broken line represents the
      power law behavior $(t-s+t_0)^{-0.80}$ 
      of $s^{1/z}R_{ag}(t,s)$ from Eq.(\ref{phen}). }
\label{fig24}
\end{figure}

Next, we have extracted $R_{ag}(t,s)$ via the
subtraction $R_{ag}(t,s)=R(t,s)-R_{st}(t-s)$ and we have made
the comparison with the fitting formula~(\ref{GAF}), which in the short time regime reads
\be
R_{ag}(t,s)= A_R s^{-1/z}(t-s+t_0)^{-1+1/z-a} \left[1+{\cal O}
\left (\frac{t-s}{s} \right) \right].
\label{phen}
\ee
and predicts TTI behavior if $s^{1/z}R_{ag}(t,s)$
is plotted against $t-s$. Indeed, this is observed in
Fig.\ref{fig24}, where we have used the exponent
$1/z=0.47$~\cite{note} obtained from the numerical data for the interface density
$\rho_I(s) \sim s^{1/z}$.  The curves for different values of $s$
collapse on a master
curve which is very well fitted by the power law
$(t-s+t_0)^{-0.80}$ (broken line in Fig.\ref{fig24}).
The comparison with Eq.~(\ref{phen}), then, gives the numerical value
\be
a=0.27 \pm 0.02 
\label{a}
\ee
in agreement with previous results for this exponent
 \cite{noiachi,noiRd2}.
The tiny deviations from the fitting curve, observed in Fig.\ref{fig24} when $s$
is small and $t-s<2$, can be attributed to the absence
of a sharp separation between bulk and interface fluctuations in this time regime.
This implies that Eq.~(\ref{splitR}) is not exact for small $s$ and, therefore, that
the aging contribution in the response function cannot be obtained  simply by 
subtracting $R_{st}(t-s)$ from $R(t,s)$. However, this procedure
becomes exact for larger values of $s$, as demonstrated by the fast
convergence of the numerical data for
$R_{ag}(t,s)$ towards the behavior of Eq.~(\ref{phen}).
Furthermore, as remarked above,
the equilibrium response $R_{st}(t-s)$ is a very fast decreasing function of
$t-s$ and, when $t-s>2$, the condition $R_{st}(t,s) \ll R(t,s)$ is
fulfilled. Hence, even for small values of $s$,  
in the time region $t-s>2$, one has always $R(t,s) \simeq R_{ag}(t,s)$ and the  
numerical curves follow Eq.~(\ref{phen}).
\begin{table}  
\begin{ruledtabular}
\begin{tabular} {|l|c|c|c|c|}
    &  $a$ &  $\theta$ & $\frac{T_CA_R}{A_C}$  & $X_\infty$  \\
  \hline
  \hline
    Gaussian model & (d-2)/2 & 0 & 1/2 & 1/2  \\
  \hline
     Ising critical $d=4$   &  $1.01 \pm 0.01$  &  $ 0.00 \pm 0.02 $ &  
       $0.49 \pm 0.03$ & $0.49 \pm 0.03$ \\
  \hline
     Ising critical $d=2$ &  0.115  &  0.38  &  $0.20 \pm 0.01$ 
     & $0.33 \pm 0.01 $ \\
  \hline
     Ising $T < T_C$ $d=2$ &  $0.27 \pm 0.02$  &    &   
     &  0  
\end{tabular}
\end{ruledtabular}
\caption{Exponents, amplitude ratio and $X_\infty$ for quenches to and to below $T_C$. The values of $a$ 
and $\theta$, in the $d=2$ critical quench of the Ising model, are taken from ref.\cite{Pleimling}. 
For $X_\infty = 0$ in the quench below $T_C$ see, for instance, ref.\cite{noiteff}}.
\label{tabella}
\end{table}

\section{Conclusions}
\label{Conclusioni}

We have investigated the suggestion  put forward in Ref.~\cite{Cala1,Pleimling}                
that LSI applies when Gaussian behavior holds, by
looking at the scaling behavior of $R(t,s)$ in the kinetic Ising model with Glauber
dynamics in two revealing
test cases i) in the quench to $T_C$ with $d=4$, where deviations from Gaussian
behavior are expected to disappear and ii) in the quench to $T_C$ and to below $T_C$ with $d=2$
where, conversely,  corrections to Gaussian
behavior are expected to become quite sizable.
Unlike Pleimling and
Gambassi, who work with an intermediate integrated response function, 
we have computed directly  $R(t,s)$, in the sense specified in section \ref{sec2} after Eq.~(\ref{trec}),
producing high precision data
by means of the new numerical algorithm of ref. \cite{noialg}. In the $d=4$ numerical simulation
we have not found logarithmic corrections to the Gaussian behavior (see the insets of
Fig.1 and Fig.2).

Our results do confirm the conjecture that a Gaussian approximation is inherent to LSI.
In the case of the $d=4$ critical quench we find agreement
between  LSI, Gaussian behavior, and numerical data. 
Instead, in the case of the critical quench in $d=2$,
deviations from LSI are observed, which go in the same direction as the
field theoretical calculations and previous numerical results from 
the global integrated response functions. 
Similarly, important deviations from LSI behavior are found in
the quench to below
$T_C$. In the latter case the data i) are incompatible with the multiplicative form
of $R(t,s)$ predicted by LSI and ii) do confirm the result $a=0.27\pm 0.02$ for the scaling
exponent of $R_{ag}(t,s)$, first obtained from the measurements of $\chi(t,t_w)$ \cite{noiachi}. 
It ought to be mentioned that the behavior of $R_{ag}(t,s)$ in the quench to below $T_C$ of the
$d=2$ Ising model has already been investigated numerically in great detail in
ref. \cite{noiRd2}. In that paper we have produced evidence for the existence 
of a strong correction to scaling, next to the leading term behaving
as in Eq.~(\ref{scal4R}).
In the present work there has been no need to worry about the correction to scaling,
since we have focused on the time
sector with $t-s<s$ and $s$ sufficiently large, where the correcting
term is negligible \cite{noiRd2}.  

Finally, it should also be mentioned that recently Henkel and collaborators~\cite{LSI.1,LSI.2} have proposed
a more general version of the LSI by replacing $F_R(x)$ in Eq.~(\ref{scalR}) with
\be
    F_R(x)=A_R\;(x-1)^{-(1+a^{\prime})}\;x^{\theta+a^{\prime}-a}
    \label{concl1}
\ee 
where $a^{\prime}$ is new exponent. The old LSI is contained in the new one as the
particular case corresponding to $a^{\prime}=a$. Fitting the numerical data for the 
integrated response function in the critical quench of the $d=2$ Ising model~\cite{LSI.2},
an improvement over the old LSI has been obtained with $a-a^{\prime}=0.187$. One of the
problems with the new LSI, however, is that when $a \neq a^{\prime}$ the numerical improvement
is obtained at the expense of destroying quasi stationarity in the short time
regime, which is required by the separation of the time scales~\cite{Cugliandolo}. 
A detailed analysis of the new LSI is
beyond the scope of the present work and is deferred to a future publication.
Considerations similar to ours have been made by Hinrichsen~\cite{Hinri} in comparing
numerical data with the predictions of LSI~\cite{contact} for the $1+1$-dimensional contact process.

\vspace{2cm}

{\bf Acknowledgments}
 
This work has been partially supported by MURST through PRIN-2004.

\newpage

\vspace{12cm}

\vskip15cm

\newpage

\vskip15cm

\vskip15cm

\newpage

\vskip15cm
\newpage

\end{document}